# A Physically Based Analytical Modeling of Threshold Voltage Control for Fully-Depleted SOI Double Gate NMOS-PMOS Flexible-FET

Nadim Chowdhury, Zubair Al Azim[1], Imtiaz Ahmed, Iftikhar Ahmad Niaz,
Md. Hasibul Alam and Dr. Quazi D.M. Khosru[2]

Department of Electrical and Electronic Engineering
Bangladesh University of Engineering and Technology,
Dhaka-1000, Bangladesh
[1]E-mail: sakifbuet@gmail.com
[2]On leave to Green University of Bangladesh, Dhaka-1207, Bangladesh

*Abstract-* In this work, we propose an explicit analytical equation to show the variation of top gate threshold voltage with respect to the JFET bottom gate voltage for a Flexible Threshold Voltage Field Effect Transistor (Flexible-FET) by solving 2-D Poisson's equation with appropriate boundary conditions, incorporating Young's parabolic approximation. The proposed model illustrates excellent match with the experimental results for both n-channel and p-channel 180nm Flexible-FETs. Threshold voltage variation with several important device parameters (oxide and silicon channel thickness, doping concentration) is observed which yields qualitative matching with results obtained from SILVACO simulations.

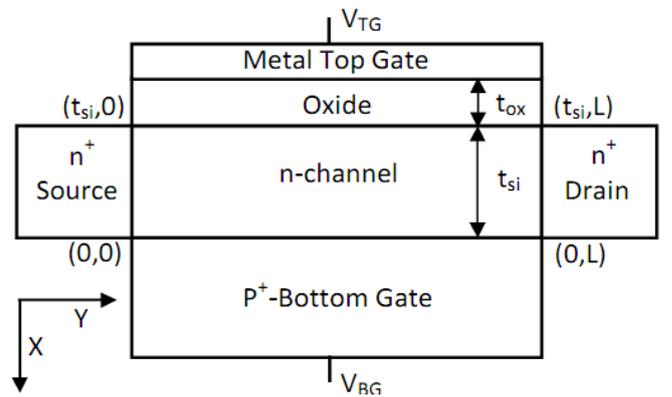

Fig. 1. n-channel Flexible-FET

## I. Introduction

Remarkable immunity to short channel effects and excellent device scaling possibilities has led to the development of double gated MOSFETs. Flexible-FET has been reported as one the most scalable[1] independently-double-gated MOSFETs with a unique dynamic threshold voltage control feature using the bottom gate voltage. Flexible-FET is the only widely reported device that combines a JFET with a MOSFET (Fig. 1). It has a damascene metal top gate and an implanted JFET bottom gate that are self aligned in a gate trench[2].

In this paper, 2-dimensional analytical modeling of dynamic threshold voltage control by the bottom gate voltage is developed using semi classical analysis. Operational physics including energy band-diagrams has also been proposed for this device. With the use of the developed model, effects of varying several key device parameters have been observed. Results from proposed model have been validated by experimental data.

## II. Device Operation

Relative doping concentrations of the n and p$^+$ side are properly chosen to ensure that the n side remains fully depleted under no external bias (Fig. 2). When a sufficiently large positive voltage is applied at the top gate, negative charges begin to accumulate at the oxide-semiconductor interface as a result of capacitive action. Charge is supplied by the drain-source regions and the oxide-semiconductor interface is effectively accumulated. Threshold condition is illustrated in Fig. 3 and accumulation is displayed in Fig. 4. If a positive voltage is applied at the bottom gate, the built-in potential barrier is reduced; so, the threshold voltage is reduced. However, if a negative voltage is applied at the bottom gate, the potential barrier is increased; so, the threshold voltage is increased.

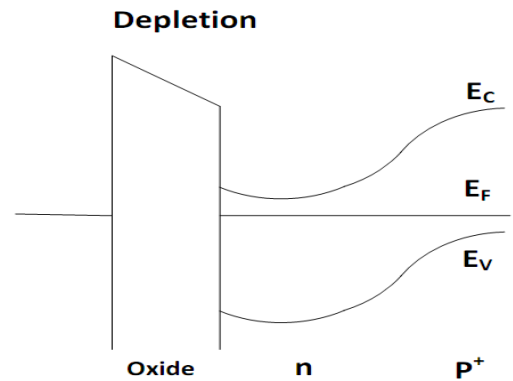

Fig. 2. Band diagram for thermal equilibrium condition

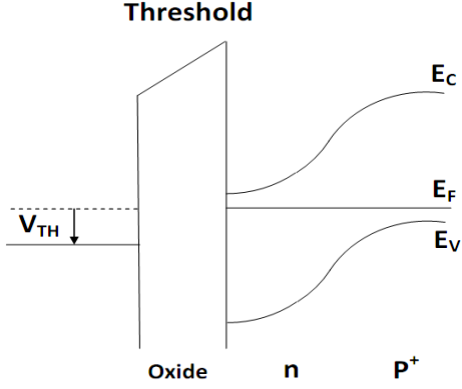

Fig. 3. Band diagram for threshold condition

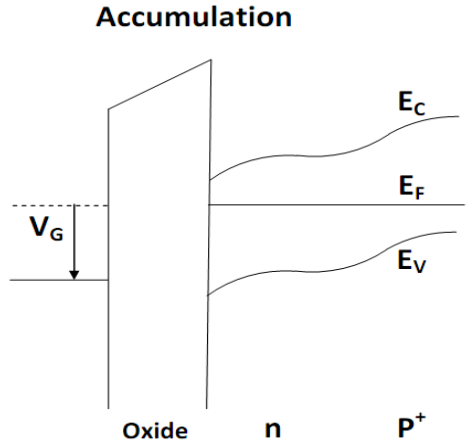

Fig. 4. Band diagram during conduction

III. ANALYTICAL MODELING

Semi classical analysis has been used for an 180nm n-channel device. The potential $\varphi(x,y)$ is assumed to vary along y-direction in accordance with Young's parabolic approximation for fully depleted SOI MOSFETs[3]:

$$\varphi(x,y) = \varphi_s(x) + C_1(x)y + C_2(x)y^2 \quad (1)$$

Where, $\varphi_s(x)$ is the surface potential and $C_1(x)$ and $C_2(x)$ are constants which are found using appropriate boundary conditions. For the continuous Electric flux at the gate-oxide interface:

$$\frac{d\varphi(x,y)}{dy}\bigg|(y=0) = \frac{\varepsilon_{ox}}{\varepsilon_{si}} \frac{\varphi_S(x) - V_{GS}'}{t_{ox}} \quad (2)$$

$$V_{GS}' = (V_{TG} - V_{FB}) \quad (3)$$

Device is subjected to following boundary conditions:
$$\varphi(L,0) = V_{bin} + V_{DS} \quad (4)$$
$$\varphi(0,0) = V_{bin} \quad (5)$$
$$\varphi(x,t_{si}) = V_{BG} - V_{bi} \quad (6)$$

$C_1(x)$ and $C_2(x)$ are determined with these conditions as:
$$C_1(x) = \frac{\varepsilon_{ox}}{\varepsilon_{si}} \frac{\varphi_S(x) - V_{GS}'}{t_{ox}} \quad (7)$$

$$C_2(x) = \frac{(V_{BG} - V_{bi}) - \varphi_S(x) - \frac{C_{ox}}{C_{si}}(\varphi_S(x) - V_{GS}')}{t_{ox}} \quad (8)$$

Poisson's equation for fully depleted channel is solved.

$$\frac{d^2\varphi(x,y)}{dx^2} + \frac{d^2\varphi(x,y)}{dy^2} = -\frac{qN_D}{\varepsilon_{si}} \quad (9)$$

$$\frac{d^2\varphi_s(x)}{dx^2} - \alpha^2 = \beta \quad (10)$$

Where,
$$\alpha^2 = 2\frac{1 + \frac{C_{ox}}{C_{si}}}{t_{si}^2} \quad (11)$$

$$\beta = -\left[\frac{qN_D}{\varepsilon_{si}} + \frac{2\frac{C_{ox}}{C_{si}}V_{GS}' + 2(V_{BG} - V_{bi})}{t_{si}^2}\right] \quad (12)$$

Surface Potential and Electric field are expressed as:
$$\varphi_S(x) = G\sinh(\alpha x) + H\cosh(\alpha x) - \frac{\beta}{\alpha^2} \quad (13)$$
$$E(x) = \alpha[G\cosh(\alpha x) + H\sinh(\alpha x)] \quad (14)$$

With,
$$G = \frac{V_{DS} + V_{bin}(1 - \cosh(\alpha L)) + \frac{\beta}{\alpha^2}(\cosh(\alpha L))}{\sinh(\alpha L)} \quad (15)$$

$$H = V_{bin} + \frac{\beta}{\alpha^2} \quad (16)$$

At the threshold point,
$$\varphi_{Smin} = \gamma\varphi_F = 2\sqrt{GH} - \frac{\beta}{\alpha^2} \quad (17)$$

Where $\gamma$ is a parameter whose value is in the range of 1-2. Solving for the threshold voltage and simplifying:

$$V_{THN} = -\frac{C_{si}}{2C_{ox}}\left\{t_{si}^2\left(\frac{qN_D}{\varepsilon_{si}} + \frac{-N + \sqrt{N^2 - 4MP}}{2M}\right) + 2(V_{BG} - V_{bi})\right\} + V_{FB} \quad (18)$$

With
$$M = 2\sinh^2(\alpha L) - 4\cosh(\alpha L)/\alpha^4 + 8/\alpha^4 \quad (19)$$
$$N = 8\varphi_F\alpha^2\sinh^2(\alpha L) + 8(V_{bi} + V_{DS})/\alpha^2 + +8/\alpha^4 - 4(2V_{bi} + V_{DS})\cosh(\alpha L)/\alpha^2 \quad (20)$$
$$P = 8\varphi_F^2\alpha^4\sinh^2(\alpha L) + 4(V_{bi} + V_{DS})^2 - 4(V_{bi} + V_{DS})V_{bi}\cosh(\alpha L) + 4V_{bi}^2 \quad (21)$$

Using (eq. (2)) the control factor ($f$) [4] is represented with the following relation:
$$f = \frac{C_{si}}{C_{ox}} = 3\frac{t_{ox}}{t_{si}} \quad (22)$$

IV. SIMULATION AND MODEL VERIFICATION

For the proposed model the surface potential variation is observed (Fig. 5). The electric field along the channel is illustrated in Fig. 6. Threshold voltage variation with Donor doping concentration (Fig. 7), Oxide thickness (Fig. 8) and Silicon thickness (Fig. 9) is observed.

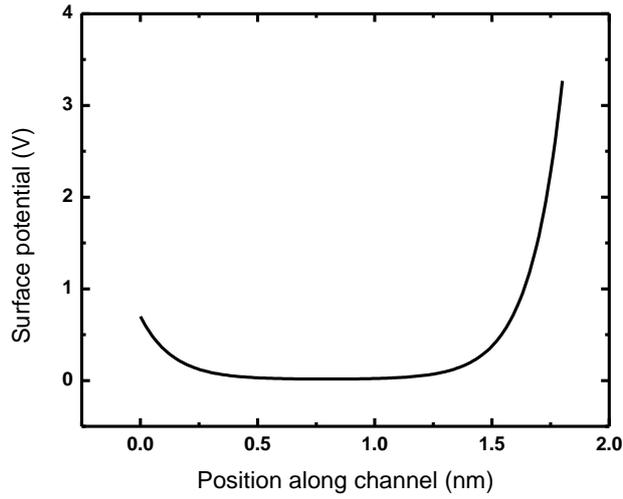

Fig. 5. Surface potential along the channel

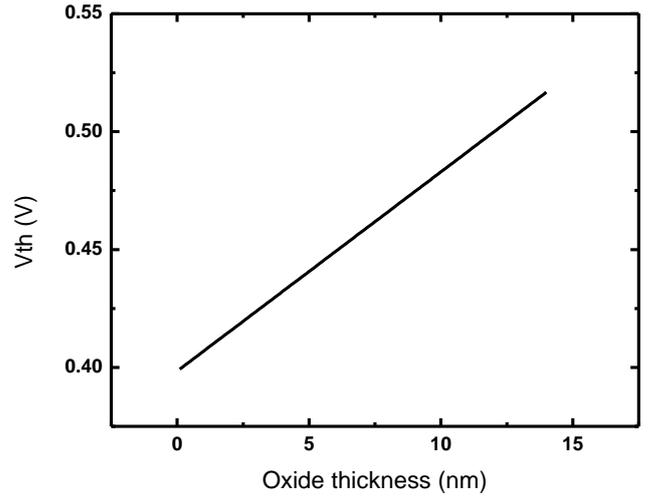

Fig. 8. Threshold voltage variation with oxide thickness

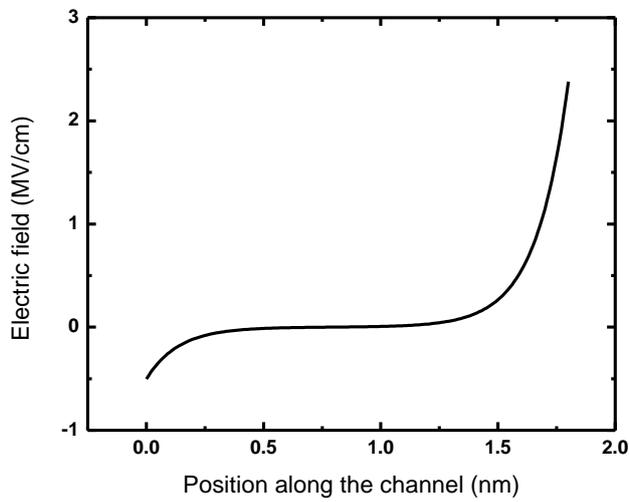

Fig. 6. Electric Field along the channel

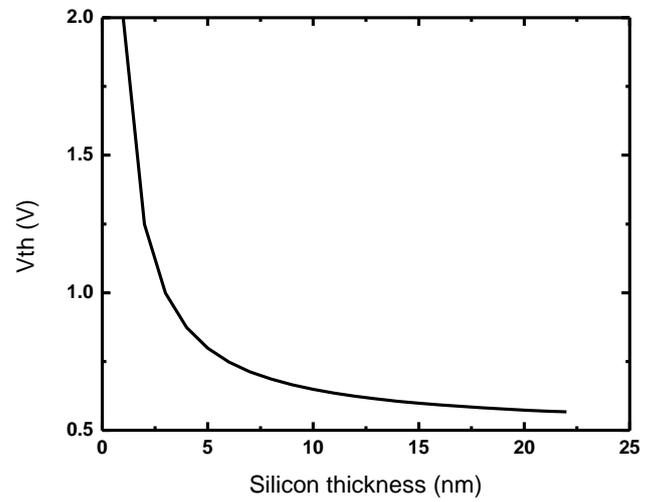

Fig. 9. Threshold voltage variation with silicon thickness

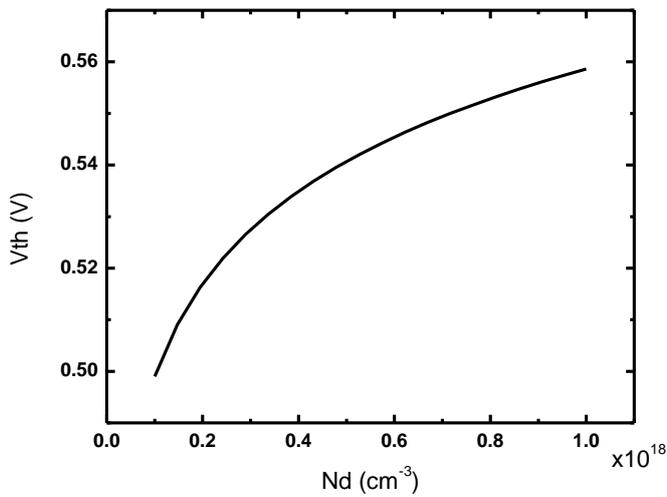

Fig. 7. Threshold voltage variation with Donor doping concentration for 180nm n-channel Flexible-FET

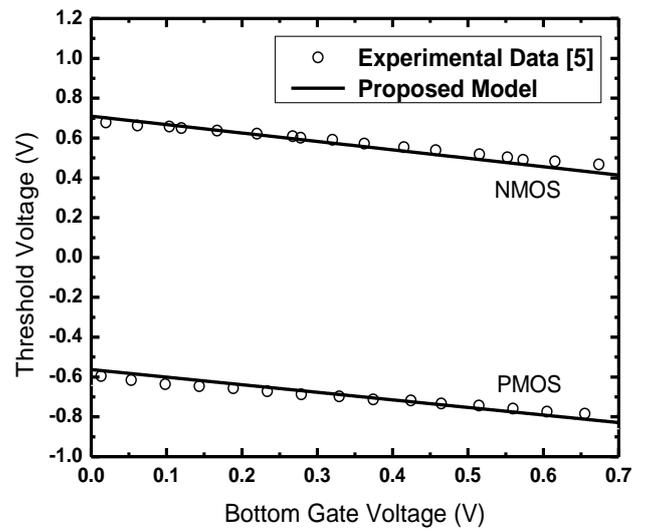

Fig. 10. Threshold voltage modeling of 180nm n-channel & p-channel Flexible-FET by Bottom gate voltage

Using the obtained analytical relation (eq. (18)), the threshold voltage variation for an 180nm Flexible-FET in accordance to bottom gate voltage has been simulated (Fig. 10) and the results are compared with experimental data [5], which shows excellent agreement. Moreover, the derived value of the control factor ($f$) (eq. (22)) agrees well with results from SILVACO [4].

## V. Conclusion

Using semi classical analysis we have been able to match the results with experimental data for both n-channel and p-channel devices. Applicability for both n-channel and p-channel devices makes the proposed model suitable for CMOS applications. CMOS Flexible-FETs are extremely suitable for ultra low power applications and our proposed analytical model can be vital in controlling their operation.